\documentclass[twocolumn,superscriptaddress,floatfix,showpacs]{revtex4-1}
\pdfoutput=1
\usepackage{graphics,amssymb,amsmath,epsfig,color,url}
\usepackage{graphicx}

\newcommand{\be}{\begin{equation}} \newcommand{\ee}{\end{equation}}
\newcommand{\bea}{\begin{eqnarray}} \newcommand{\eea}{\end{eqnarray}}
\begin{document}

\title{Polymer collapse and crystallization in bond fluctuation models}
\author{Peter Grassberger} \affiliation{JSC, FZ J\"ulich, D-52425 J\"ulich, Germany}
         \affiliation{Max Planck Institute for the Physics of Complex Systems, N\"othnitzer Strasse 38, D-01187 Dresden, Germany}
\date{\today}

\begin{abstract}

While the $\Theta$-collapse of single long polymers in bad solvents
is usually a continuous (tri-critical) phase transition, there are exceptions where it is preempted by 
a discontinuous crystallization (liquid $\leftrightarrow$ solid) transition. For a version of the 
bond-fluctuation model (a model where monomers are represented as $2\times 2\times 2$ cubes, and 
bonds can have lengths between 2 and $\sqrt{10}$) it was recently shown by F. Rampf {\it et al.} 
that there exist distinct collapse and crystallization transitions for long but {\it finite} chains. 
But as the chain length goes to infinity, both transition temperatures converge to the same $T^*$, 
i.e. infinitely long polymers collapse immediately into a solid state. We explain this by the 
observation that polymers crystallize in the Rampf {\it et al.} model into a non-trivial cubic 
crystal structure (the `A15' or `Cr$_3$Si' Frank-Kasper structure) which has many degenerate ground states and, 
as a consequence, Bloch walls. If one controlls the polymer growth such that only one ground state 
is populated and Bloch walls are completely avoided, the liquid-solid transition is a smooth 
cross-over without any sharp transition at all.

\end{abstract}

\maketitle


In spite of having been studied for several decades, the phase transitions of a single long polymer 
in a bad solvent are still a subject of active research with occasional big surprises. When temperature
is lowered, most polymers 
undergo a $\Theta$-collapse, which is typically a continuous (tri-critical) phase transition. At 
the transition point, repulsive entropic effects and attractive energetic forces 
would cancel exactly for an infinitely long chain. For finite chain length $N$
theory \cite{duplantier1987,hager1999theta} predicts logarithmic corrections which are also seen in 
simulations \cite{grassberger1995Theta,grassberger1997PERM,hager1999theta} and in experiments
\cite{boothroyd1992}, and which also affect the unmixing transition of long polymers 
\cite{frauenkron1997}. Qualitatively, the $\Theta$-transition resembles a gas-liquid transition,
with the open coil resembling the gas and the dense globule being analogous to a liquid. At even 
worse solvent conditions, in many models (e.g. in off-lattice polymers with Lennard-Jones 
interactions between monomers \cite{Schnabel09}) there occurs a second `liquid-solid' transition 
that is similar to crystallization. Since it is a discontinuous transition, its properties are 
usually not universal and are not described by perturbative renormalization group methods.

But there are well known cases where this $\Theta$-collapse is preempted by a discontinuous
`freezing' or crystallization transition. One example is semi-stiff polymers \cite{bastolla1997},
another is provided by (off-lattice) monomers with hard core repulsion and attractive interactions
which extend only very little beyond the core \cite{taylor2009phase,taylor2009all}. Still another is 
the cooperative one-step collapse of some proteins \footnote{
Notice that only monomer positions, not the the location of bonds is frozen out in
homopolymer crystallization, while also the latter are frozen out in protein folding.
Thus the often cited analogy between protein folding and the crystallization of homopolymers 
should not be stretched too far, as the cooperativity of protein folding is entirely due to their
heterogeneity.}

A very surprising behavior which falls between these two possibilities -- $\Theta$-collapse with 
subsequent liquid-solid transition, and immediate gas-solid freezing -- was observed by 
Rampf {\it et al.} \cite{Rampf05,rampf2006phase,paul2007unexpectedly,binder2006simulation,binder2008phase}.
In their model, the specific heats of finite chains show rounded
peaks at distinct temperatures $T_{\rm crys}(N) < T_\Theta(N)$, but as the chain length $N$ 
diverges, 
\be
   \lim_{N\to\infty} T_{\rm crys}(N) = \lim_{N\to\infty} T_\Theta(N) = T^*.
\ee
Thus, while finite chains show the expected two distinct transitions, infinitely long polymers
collpase in a single ``hybrid" transition into their ground state \footnote{Hybrid transitions
that combine aspects of both first and second order have recently been found also in other models
\cite{causo2000,dorogovtsev2006,schwarz2006,goltsev2010,bizhani2012}, but there seems to be 
no closer connection.}.

In \cite{taylor2009phase,taylor2009all} it was conjectured that this is connected to the 
range of interactions in this model, since the phenomenon 
disappeared when also more distant monomer pairs are included in the interaction. But it 
is not clear what would be the detailed mechanism for that. 
 In the present paper, we propose a 
different explanation which is based on the fact that discontinuous phase transitions arise 
typically if there is a large gap between the (few) low energy states that dominate the low-$T$ 
phase and the bulk of states at higher energies. This is of course also the reason why there 
is a first order collapse in hard-core polymer models with very short range attraction:
The shorter the range of the attractive potential, be bigger will be the effect of a single 
displaced monomer, as it will cause many monomer distances to be out of the interaction range.
In the Rampf et al. model, the energetic distance between ground and excited states is, 
in contrast, dominated by topological defects similar to Bloch walls in magnetic systems
that arise from a spontaneously broken symmetry.

We verify this conjecture by studying also a modified model where all such topological 
defects are completely eliminated, by constraining all configurations to be in the same 
ergodic component. This modified model has of course the same ground state energy, but 
it has a dramatically changed phase diagram: There is a standard $\Theta$ transition, but 
no crystallization transition at all. Instead, the transition from a disordered ``liquid" to 
the densely packed ground state is a continuous cross-over. A similar behavior seems to 
prevail for interacting self-avoiding walks on the simple cubic lattice, although the 
interpretation there is controversial \cite{vogel2007}. In the very careful studies 
of \cite{vogel2007}, clear peaks in the specific heat were observed which obviously
are related to {\it some} crystallization phenomena. But all indications point to the fact
that they are not associated with bulk crystallization, but with changes in the
geometric shapes and surface properties of clusters which are already completely compact 
and ordered in their bulk.



The model used by Rampf {\it et al.} is the bond fluctuation model (BFM) of Carmesin {\it 
et al.} \cite{carmesin1988}, augmented by an ansatz for the interactions. The BFM lives on
a simple cubic lattice. Monomers are represented by $2\times 2\times 2$ cubes, i.e. if site 
${\bf x} = (x,y,z)$ is occupied by a monomer, all sites with Euclidean distance $< 2$ are excluded.
These monomers are linked by bonds with integer coordinate components and length $2 \leq l \leq \sqrt{10}$.
Thus the set of all bond vectors is made up by the vectors $(2,0,0), (2,1,0), (2,1,1), (2,2,0), (2,2,1), (3,0,0),
(3,1,0)$ and all their rotations and reflections (altogether 108 vectors).

The solvent is not treated explicitly, but implicitly by assuming an attractive (negative) contact 
energy for any pair of monomers that touches each other along part of their surface (line and corner
contacts do not contribute to this energy). Notice that two monomers can touch on their full $2\times 2$
sides, or on contact areas of size $2\times 1$ or $1\times 1$. If every contact would make a contribution
$-\epsilon {\cal A}$ to the total energy (where $1 \leq {\cal A} \leq 4$ is the contact area), the 
total energy of compact configuration (i.e., of a tiling of 3-d space by $2\times 2\times 2$ cubes) 
would be $-12\epsilon$ per monomer, independent of the way how space is tiled.

This is, however, not the interaction model chosen in \cite{Rampf05}. There it was assumed that 
every touching pair (including those which are joined by a bond) contributes $-\epsilon$ to the 
energy, independent of the area of contact. Thus 
for achieving minimal energy it is preferred to avoid full-side contacts and to maximize the number 
of surfaces which are in partial contact. In the following we shall assume that the minimal 
energy configuration is indeed a tiling, i.e. the monomers occupy space densely. We have no 
rigorous proof for this, but it is fully compatible with the simulations. As shown in \cite{Sikiric},
there exist nine inequivalent tilings of space by cubes of the same size. All of them except one
are made up by perfect layers. The one not made up by perfect layers is the one with the largest 
number of contacts. It is periodic, and every cube in it has in average 27/2 neighbors in contact
(the next best tiling is by layers consisting of simple square lattices, each layer shifted by 
by one unit. In this tiling, each cube has 12 neighbors).

Indeed, this optimal tiling is nothing but the A15 (or Cr$_3$Si) Fank-Kasper structure 
\cite{deGraef2012}. Usually, in this structure particles are represented not by cubes but by 
spheres, and their centers are then slightly displaced with respect to the ideal lattice 
positions. In nature, this crystal structure is realized for binary alloys of composition 
A$_3$B where the A atoms are slightly larger than the B atoms (as far as I am aware of, the 
fact that the A15 structure corresponds to a cube tiling was not known before). All alloys with 
A15 structure are very brittle, due to the absence of easy gliding planes when seen as packings 
of cubes. The A atoms have coordination number 14, while the B atoms have coordination number 12,
giving rise to the average 27/2 mentione above. Seen as a cubic crystal structure, the basic 
unit cell has size $4\times 4\times 4$ and contains 8 atoms (monomers) at the sites
\bea
  && (0,0,0),\; (1,1,2),\; (1,2,0),\; (1,3,2),\; \nonumber \\
  && (2,0,0),\; (3,0,2),\; (3,2,1),\; (3,2,3). \label{sites}
\eea 
For any perfect arrangement
of atoms in such a structure there exist many other arrangements which are just shifted copies of it. 
Thus this structure, seen as a ground state of a statistical mechanics model, is highly degenerate.
If we prepare a big system such that it is in one of these ground states in some region of 
space of characteristic size $L$ and in another ground state somewhere else, there must be 
``Bloch walls" between them. These are toplological defects with energies $\propto L^2$. It is these 
Bloch walls which are responsible for the discontinuity and high temperature of the freezing 
transition.   


In contrast to \cite{Rampf05,rampf2006phase,paul2007unexpectedly,binder2006simulation,binder2008phase},
where Markov chain Monte Carlo methods were used, we used PERM \cite{grassberger1997PERM} and 
``flat PERM" \cite{prellberg2004flat}.
PERM is a growth algorithm with importance sampling and re-sampling, i.e. chains are grown
in a biased way, with the bias compensated by weight factors. On the basis of weights of partially 
grown chains, the ``population" of chains is controlled, similarly as in genetic models, by 
pruning and cloning. In the original PERM algorithm the growth is controlled by comparing the weight
of the current configuration with the estimated partition sum, i.e. with the average weight of 
all previous configurations with the same length. In flat PERM it is compared with the average weight 
of those previous configurations that have the same length {\it and energy}. Both versions of the 
algorithm work perfectly, if the current weights of partially grown chains
are good indicators of their final weights when the growth is finished. This is true for 
polymers at temperatures above $T_\Theta$, and even more so at $T_\Theta$, where its success
is spectacular \cite{grassberger1997PERM}. But it deteriorates quickly, if growth conditions 
change much during the growth and configurations which finally would be very good are pruned away
at early stages. This happens typically in first order phase transitions. While Markov chain Monte 
Carlo methods are based on the concept of changing complete configurations to increase their 
Boltzmann weight, in PERM no configuration can be changed once it has been created.

\begin{figure}
\includegraphics[width=0.52\textwidth]{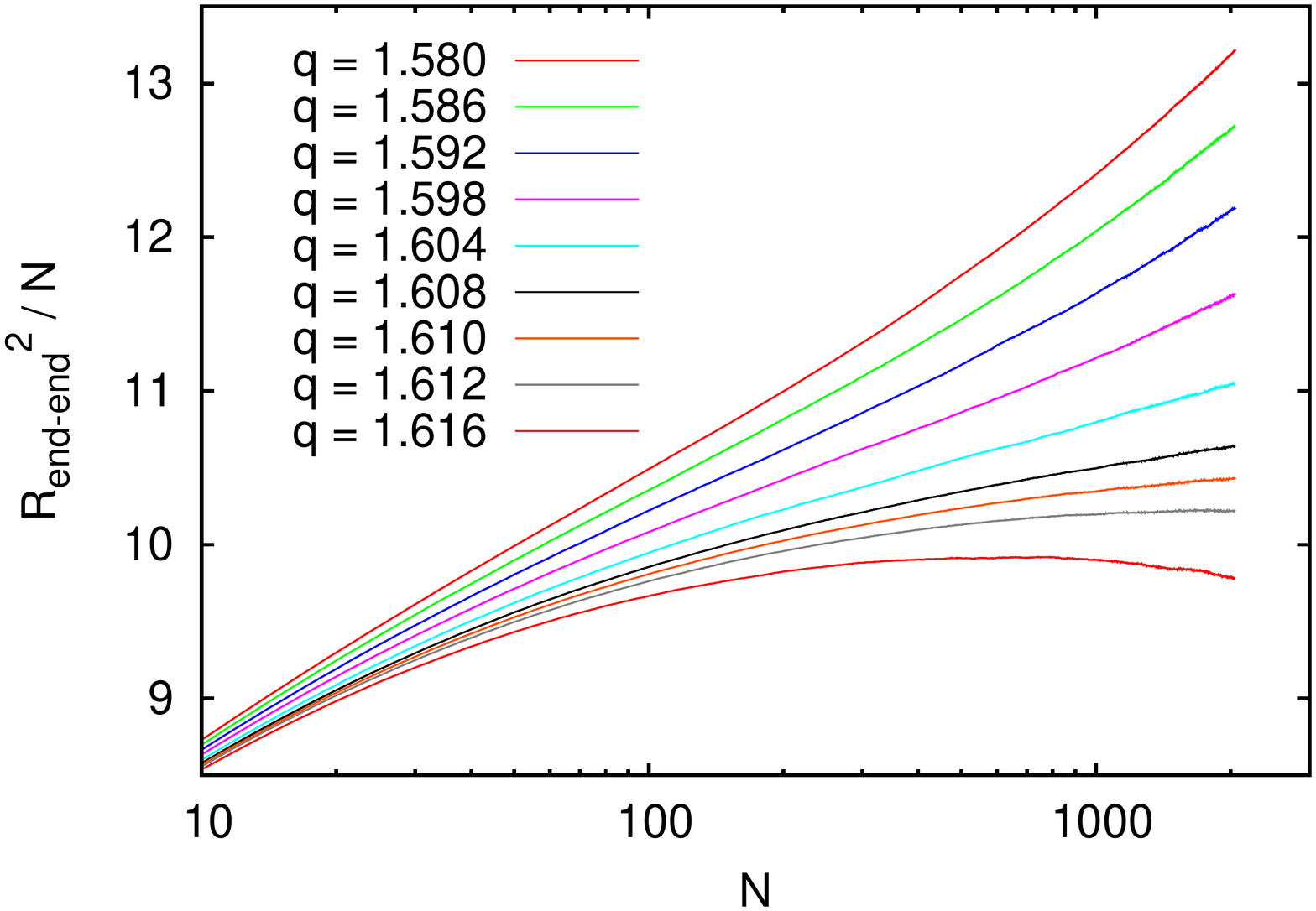}
\caption{(Color online) Semi-logarithmic plot of $N^{-1}$ times the average squared end-to-end distances against 
chain length $N$. Instead of temperature, we use the Botzmann factor $q = \exp(\epsilon/k_BT)$ to 
label each curve. At $T=T_\Theta$ one expects a horizontal curve up to logarithmic corrections.}
   \label{R2.fig}
\end{figure}

In addition, we tried also another version called ``new PERM" (nPERM) \cite{hsu2003growth}. All 
three methods gave very similar results. 
 Typical ``good" results are shown in Fig.~1. There we present estimates of the average squared end-to-end
distances, plotted against $N$ for several temperatures near $T_\Theta$. At $T=T_\Theta$ one 
expects 
\be
   R^2_{\rm end-end}\propto N,     \label{R2}
\ee
 up to logarithmic corrections. Although the leading corrections
are known theoretically \cite{duplantier1987,hager1999theta}, it is well known that they do not 
quantitatively describe the behavior at presently reachable chain lengths 
\cite{grassberger1995Theta,grassberger1997PERM,hager1999theta}. Nevertheless we clearly see 
deviations from Eq.~(\ref{R2}) which qualitatively agree with the theoretical predictions. We fully 
confirm the estimates for $T_\Theta$ given in 
\cite{Rampf05,rampf2006phase,paul2007unexpectedly,binder2006simulation,binder2008phase}, but 
due to the much longer chains and higher statisitics we can give much smaller error bars in spite
of the theoretical uncertainty:
\be
    T_\Theta = 2.103(10).
\ee

\begin{figure}
\includegraphics[width=0.52\textwidth]{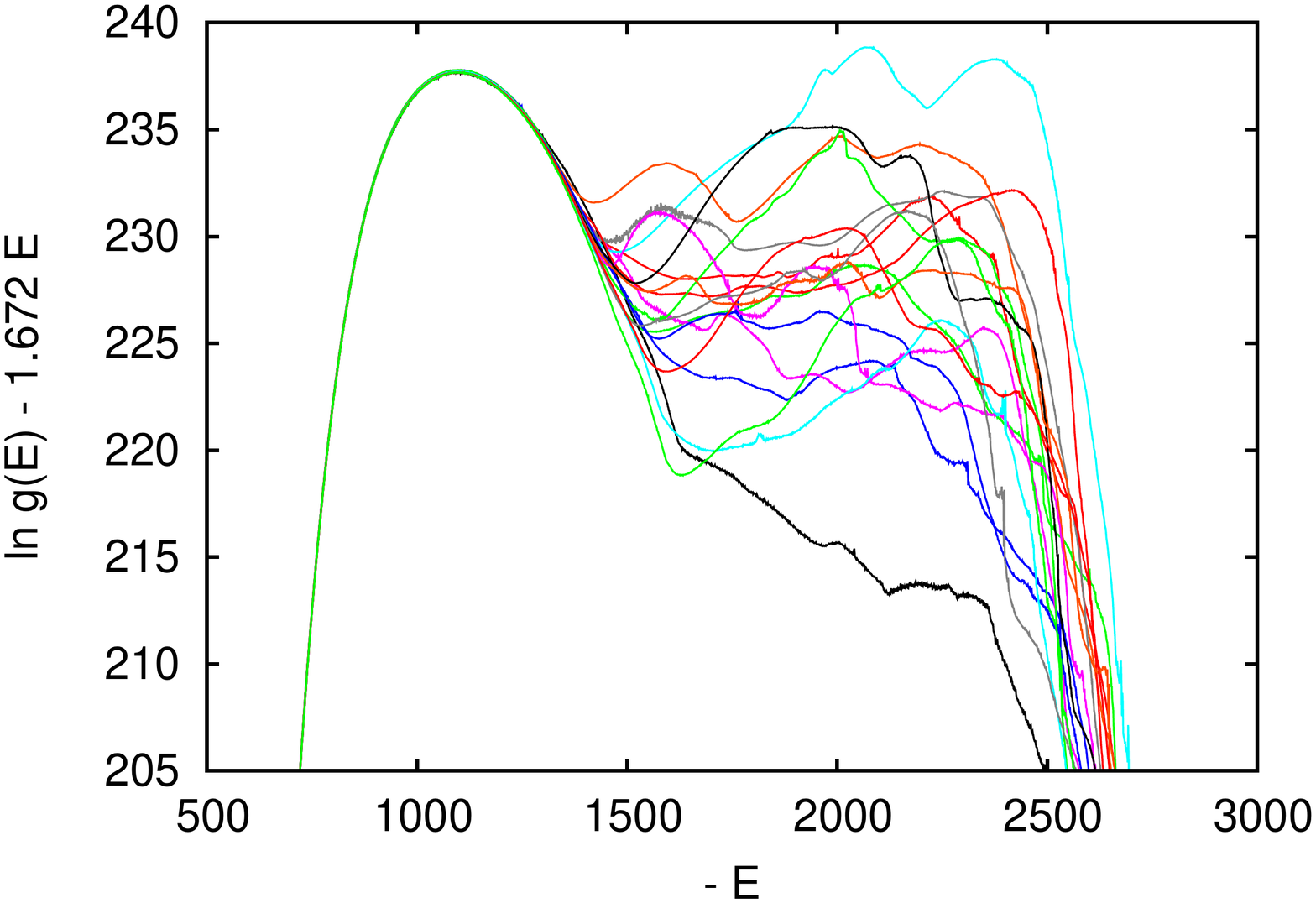}
\caption{(Color online) Logarithms of the density of states $g(E)$, plotted against tne energy $E$. In 
order to see more clearly the two humps expected from the fact that there is a first order freezing 
transition, we actually plotted $\log g(E) - 1.672 E$. Each curve is from a different flat PERM run 
with different random number seeds and using slightly different versions of the algorithm, thus ideally 
these curves should collapse.}
   \label{gE-512.fig}
\end{figure}

One advantage of PERM is that it allows to estimate {\it absolute} values of partition sums
and of the density of states $g(E)$, where $E$ is the total energy. Using $\epsilon=1$ in the 
following, $|E|$ is just the number of contacts. Indeed, these estimates are needed to controll the 
population growth, so estimating them is an essential part of the algorithm. In Fig.~2
we show $\ln g(E) + 1.672 E$ at fixed $N=512$ against $E$ for several runs. Notice that 1.672 is 
the value of the Boltzmann factor per contact at the effective freezing temperature according to \cite{Rampf05}.
Thus we expect the curve to be double-humped with both maxima having roughly the same height.
Each run used 
different random number generator seeds and different minor parameters in the algorithm, 
and ran for about 
one week of CPU time on a fast PC. We see very good results for large $E$ (i.e. for energies 
relevant at $T\geq T_\Theta$), but a complete failure at energies which, according to \cite{Rampf05},
dominate the crystallized phase. The only positive statement these simulations allow to draw 
is that there is presumably really a first order transition, since otherwise the failure 
would not be expected to be as dramatic. Similar plots were obtained for all $128 \leq N \leq 1024$.

\begin{figure}
\includegraphics[width=0.52\textwidth]{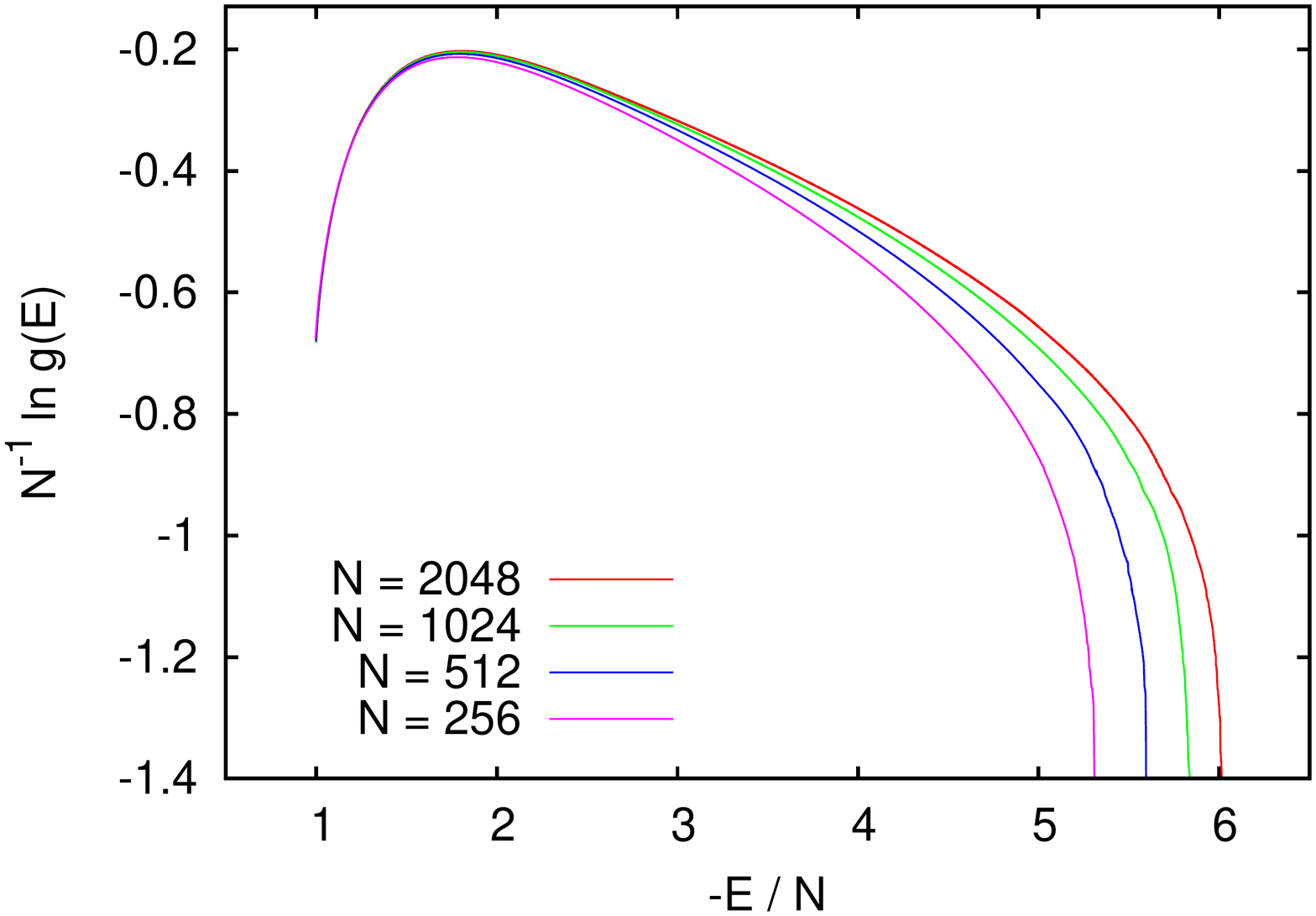}
\caption{(Color online) Logarithms of the density of states $g(E)$ per monomer, plotted against the 
energy per monomer, for the modified model restricted to the sites in a perfect crystal.}
   \label{gE-constraint.fig}
\end{figure}

\begin{figure}
\includegraphics[width=0.52\textwidth]{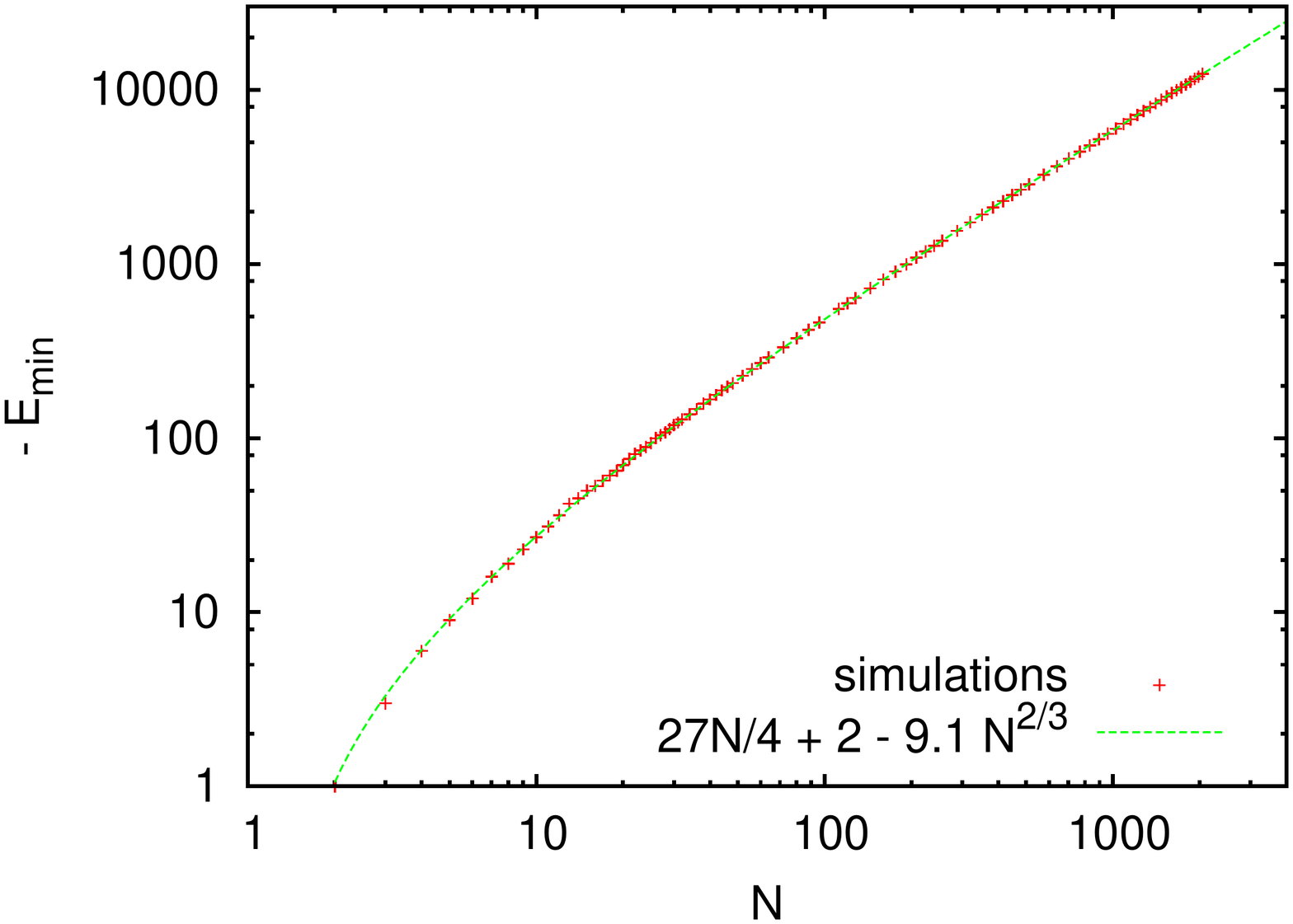}
\caption{(Color online) Energy minima reached in simulations of the modified model for $N=2048$.
The continuous curve is a fit with the predicted bulk term $27/4 N$ and a surface term $\propto N^{2/3}$.}
   \label{emin.fig}
\end{figure}

Let us now finally look at simulations of the constrained model, where we allowed monomers 
to be placed only at one of the positions given in Eq.~(\ref{sites}) (Fig.~3). This time we show data 
for different values of $N$, up to $N=2048$. The data now look absolutely clean. Moreover,
we obtained good data down to energies which are close to the expected ground state energy
$E_{\rm min} = -6.75 N + O(N^{2/3}$ (see Fig.~4). It could of course be that these results are spurious, 
and that the simulations just missed those configurations which would have created troubles. 
We consider this as highly unlikely and take Fig.~3 as an indication that there is really 
no crystallization transition. Instead, when temperature is lowered, holes in the configuration
are gradually filled up until a compact configuration is reached. Indeed, Fig.~3 looks much 
better than analogous plots for the simple cubic lattice. There we of course also have no
toplogocal defects, but the much lower coordination number implies that the growth is 
much more often blocked at low temperatures. For the fcc lattice, however, where coordination number 
is 12, we again found very clean results.

In summary, we have shown that the very strong discontinuous crystallization transition in 
the polymer model of \cite{Rampf05} is due to spontaneously broken translational symmetry in 
its low energy state. Similar discontinuous transitions are expected (and found) in continuum
models, if the polymer can freeze into a periodic monomer configuration. We do not expect any 
first order transition if the low energy state is glassy (such as in a Lennard-Jones homopolymer 
with bond lengths incommensurate with the Lennard-Jones radius) and in lattice models without 
spontaneous symmetry breaking, such as e.g. the simple cubic lattice. The reason why crystallization 
is qualitatively different in the bond fluctuation model of \cite{taylor2009all,taylor2009phase} is 
primarily not because of the increased average range of interactions, but because of the changed 
structure of the ground state and low energy excitiations.

For very helpful correspondence I thank W. Paul, P. Paufler, H. Chat\'e and M. Dutour Sikiric.
\bibliography{mm}

\end{document}